%% file: family_Aug27.tex
\documentclass[12pt,reqno]{amsart}
\usepackage{graphicx}
\usepackage{epstopdf}
\usepackage{amssymb,amscd,amsbsy}
\usepackage{amsthm}
\input{definitions}

\theoremstyle{remark}

\newtheorem*{rem*}{Remark}

\begin{document}

\newcommand{\vse}{\vspace{.2in}}
\numberwithin{equation}{section}

\title{ The family of analytic  Poisson brackets for the Camassa--Holm hierarchy.}
\author{M. I. Gekhtman and K. L.  Vaninsky}

\begin{abstract}
We consider the integrable Camassa--Holm  hierarchy on the line with positive initial data  rapidly decaying at infinity.    
It is known that  flows of the  hierarchy can be formulated in  a Hamiltonian form using two compatible Poisson brackets. In this note we propose a new approach to  Hamiltonian theory of the CH equation. In terms of associated Riemann surface and the Weyl function we write an analytic formula  which produces a family of compatible Poisson brackets. The formula includes  an entire  function  $f(z)$ as a parameter. The simplest choice  $f(z)=1$ or $f(z)=z$ corresponds  to the rational or  trigonometric solutions of the Yang-Baxter equation and produces two original Poisson brackets.  All other Poisson brackets corresponding to other choices of 
the function $f(z)$  are new. 
\end{abstract}
\maketitle
\tableofcontents 

\setcounter{section}{0}
\setcounter{equation}{0}

\section{Introduction.}
\subsection{ The Camassa--Holm  hierarchy}
The Camassa--Holm equation, \cite{CHH, CH},
\beq\label{chh}
{\p v\over \p t} + v{\p v\over \p x} +{\p  \over \p x}R\[v^2 +{1\over 2} \({\p v\over \p x}\)^2\]=0
\eeq
in which $t\geq 0$ and $  -\infty < x < \infty$, $v=v(x,t)$ is velocity, and $R$ is inverse to $L=1-d^2/d x^2$, {\it i.e.}
$$
R[ f] (x)= {1\over 2} \int\limits_{-\infty}^{+\infty} e^{-|x-y|} f(y) dy,
$$
is an approximation to the Euler equation describing  an ideal fluid. 
Introducing the function $m=L[v]$ one  writes the equation in the form
\footnote{$\;$We use   notation $D$ for the $x$-derivative and  $\bullet$ for the $t$--derivative. We use $\delta$ for the Frechet derivative.}
$$
m^{\bullet}+\(mD+Dm\)v=0. 
$$
The CH equation is a Hamiltonian system 
$m^{\bullet} +\{m,\HH\}_{J_0}=0
$
with  Hamiltonian
$$
\HH={1\over 2}\int_{-\infty}^{+\infty} mv\, dx=\text{energy}
$$
and the bracket
\beq\label{ppb}
\{A,B\}_{J_0}=\int_{-\infty}^{+\infty}{\delta A\over \delta m} \,J_0\, {\delta B\over \delta m}\, dx,\qquad\qquad J_0=mD+Dm.
\eeq

We consider the CH equation with nonnegative ($m\geq 0$) initial data and such   decay at infinity that: 
\beq\label{dec}
\int_{-\infty}^{+\infty} m(x) e^{|x|} dx < \infty.
\eeq
We denote this class of functions by $\MM$.
For such data  a solution of the initial value problem  exists for all times, see \cite{CM}.

\noi
The CH Hamiltonian is  one of infinitely many conserved integrals of motion
\bey
\HH_{0} &=& \int _{-\infty}^{\infty}\sqrt{m}\, dx,\\
\HH_{1}&=&\int_{-\infty}^{\infty} v\,dx,\\
\HH_{2}&=&\frac{1}{2}\int_{-\infty}^{\infty} v^2+(Dv)^2\,dx,\\
\HH_{3}&=&2\int_{-\infty}^{\infty} v\[v^2+(Dv)^2\]\,dx,\\
\HH_{4}&=&{1\over 2} \int_{-\infty}^{\infty}v^4\,dx+\int_{-\infty}^{\infty}v^2(Dv)^2\,dx+2\int_{-\infty}^{\infty}\[v^2+{{(Dv)^2}\over 2}\]G\[v^2+{(Dv)^2\over 2}\]\,dx,\; etc.
\eey
The quantity  $\HH_0$ is a Casimir of the bracket. 
The    integral 
$$
\HH_{1}=\int_{-\infty}^{+\infty} m\, dx=\text{momentum}
$$
produces the flow of translation $$m^{\bullet}+\{m,\HH_{1}\}_{J_0}=m^{\bullet}+Dm=0.$$
The second is the  CH Hamiltonian $\HH_{2}=\HH$.   
The  integrals $\HH_{3},\; \HH_{4},\; etc.,$ produce  higher  flows of the CH hierarchy.

Let us define the second Poisson bracket  
\beq\label{spb}
\{A,B\}_{J_1}=\int_{-\infty}^{+\infty}{\delta A\over \delta m} \,J_1\, {\delta B\over \delta m}\, dx,\qquad\qquad J_1=D-D^3.
\eeq
The conserved quantity $\HH_{1}$ is a Casimir of this bracket. The Hamiltonians of the CH hierarchy satisfy the reccurence relation 
\beq\label{recc}
J_0 \frac{\delta \HH_n}{\delta m} = J_1 \frac{\delta \HH_{n+1}}{\delta m},\qquad\qquad n  \geq  0.
\eeq

\noi
For integration of  the CH equation we consider an  auxiliary string spectral problem, 
\footnote{ $\;\;$  We use prime $'$ to denote $\xi$-derivative.}
$$
f''(\xi)+ \l g(\xi)f(\xi)=0,\qquad\qquad -2\leq \xi \leq 2.
$$
The background information for  this spectral problem can be found in  \cite{DM, GK, KK2}. 
The variables $\xi$ and $x$ are related by 
$$
x\longrightarrow \xi =2\tanh {x\over 2}.
$$
Also the potential $g(\xi)$ is related to $m(x)$ by the formula $g(\xi)=m(x)\cosh^4{x\over 2}$. 
For initial data from $\MM$ the total mass of  associated string is finite $\int_{-2}^{+2} g(\xi)d\xi <\infty$.

Two important solutions $\varphi(\xi,\l)$ and $\psi(\xi,\l)$ of the string spectral problem  are specified by   initial data
\bey
\label{init_val}
\varphi (-2,\l)& =1\qquad\qquad\qquad \;   \psi(-2,\l)=0\\
\varphi'(-2,\l)&=0\qquad\qquad\qquad      \psi'(-2,\l)=1. 
\eey 
The  Weyl function is defined by the formula 
$$
E_0(\l)=-\frac{\varphi(2,\l)}{\psi(2,\l)}.  
$$ 
The Riemann surface $\Gamma$ associated with the string spectral problem consists of two components, 
$\Gamma_+$ and $\Gamma_-$, which are  two copies of the Riemann sphere. The points $\l$'s where two spheres are glued to each other are  points of the Dirichlet spectrum. The points $\gamma$'s on $\Gamma_-$ are the points of the Newmann spectrum, see Figure 1.

\begin{figure}[ht]
\includegraphics[width=0.50\textwidth]{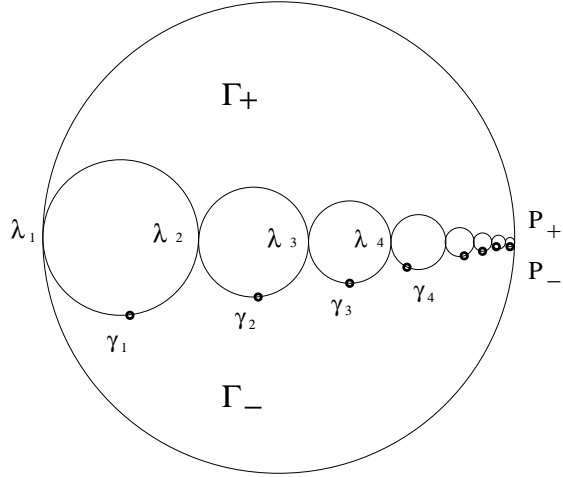}
\caption{The  Riemann surface.}
\end{figure}

The pair 
$(\Gamma, E_0)$ provides a parametrization of the phase space $\MM$, see \cite{V0}. 

\noi
Now we introduce a family of compatible Poisson brackets. 
Changing  spectral variable $\l \rightarrow z=-1/\l$ we have
$$
E_0(z)=-\frac{1}{4} +\sum_{m=1}^{\infty} \frac{\rho_m'}{z_m-z}, 
$$
where  poles $z_m$ accumulate near the origin. 
Consider a differential $\omega_{p \, q}^{f}$ on the lower component of the spectral curve $\Gamma_-$ which depends on the  entire function $f(z)$ and two points $p$ and $q$ 
$$
\omega_{p\, q}^f=\frac{1}{2\pi i}\, \frac{f(z) \, \, d z}{(z-p )\, (z-q)}\times  E_0(z) \(E_0(p)-E_0(q)\). 
$$
Any point  $p \in \Gamma_-$ can be viewed as a  function on pairs,  $p: (\Gamma, E_0 ) \rightarrow E_0(p)$.   
For any two points $p$ and $q$ that are away from the poles of some  $E_0$ we define the Poisson bracket in   
vicinity of the pair $(\Gamma, E_0)$  by the formula 
\bey
\{E_0(p), E_0(q)\}_{\omega^f}\equiv \sum_{m}\;\,  \int\limits_{\overset{\curvearrowright}{O}_m} \omega_{p\, q}^f, 
\eey
where the circles $O_m$ enclose points $z_m$ and  are traversed clockwise. The map $(\Gamma, E_0)\longrightarrow \MM $ induces  a Poisson structure on $\MM$.

It turns out that if $f(z)=1$, then the Poisson bracket defined by this formula coincides with the  bracket (\ref{ppb}). When $f(z)=z$  the Poisson bracket  coincides with (\ref{spb}). All other brackets corresponding to other choices of $f(z)$ are new, though we do not know their 
explicit form on the phase space $\MM$.  

The formula for $\omega_{p \, q}^{f}$ 
has a clear geometrical meaning. The first piece is a differential of the third kind with poles at the points $p$ and $q$. It also can have a pole at infinity depending on the rate of growth of the function $f(z)$. The second piece is quadratic in the function $E_0$ and skew-symmetric in $p$ and $q$. The formula has an immediate  generalization to  higher genus  curves, which we will discuss in future publications.

\subsection{Historical remarks.} V.I. Arnold established a Hamiltonian nature of  Euler equations of hydrodynamics using infinite dimensional diffemorphism group, 
see \cite{A1,A2}.  Later his  Lie group approach was employed by Khesin and Misiolek \cite{KM} for the CH equation. They considered Virasoro-Bott group of diffeomorphisms of the circle and introduced some special metric on it. We want to note that within this geometric approach two Poisson structures for the CH equation are obtained  in two different ways.  Within our approach employing parametrisation $(\Gamma, E_0)$ of the phase space all these compatible Poisson structures are written in a simple unified way.   

It is worth mentioning that  for two special choices of $f(z)=1$ or $f(z)=z$ our analytic formula leads to two specific solutions of the classical Yang-Baxter equation. These  rational and trigonometric solutions  determine structural constants of   a quadratic Poisson algebra \cite{GF}.   Thus our analytic approach
yields Poisson structures traditionally obtained within a group-theoretic framework
(see e.g. \cite{OPRS}).


Our formula for  Poisson bracket in terms of contour integration of some special differential is a further development of ideas of \cite{FG}. In the case $f(z)=1$ it leads to the formula for the Atiyah-Hitchin  bracket on  Weyl function, \cite{AH,V1}.

\section{Analytic Poisson brackets on  meromorphic functions.}   
In this section we introduce an analytic Poisson bracket and study its properties. 
We consider our construction in the simplest setting of rational functions on the Riemann sphere. Obvious modifications will be made for the case 
of Weyl (meromorphic) functions associated with the string spectral problem.
\subsection{Definition of the bracket.} 
\noi
We consider the space of pairs $(\CP, \chi)$,  where $\CP$ is a Riemann sphere with one marked point and $\chi(z)$ is a rational function of degree $N$. Marked point is called infinity $(\infty)$  and it corresponds to the value $\infty$ of the fixed unformization parameter $z$. The function  $\chi(z)$ is a rational function of degree $N$ which vanish at infinity.  Every  such  function can be written in the form 
$$
\chi(z)= \frac{\S(z)}{\D(z)}= \sum_{k=1}^{N}\chi_k(z) = \sum_{k=1}^{N} \frac{\rho_k}{z_k-z},
$$
where $\S(x)$ is of degree $N-1$ and $\D(z)$ is monic of degree $N$ with distinct roots; $\Rat_N$ is the space of all such rational functions.  

Now we are ready to introduce a family of compatible Poisson brackets on the space of pairs. 
Consider a differential $\omega_{p \, q}^{f}$ which depends on the  entire function $f$ and two points $p$ and $q$ on the curve  
$$
\omega_{p\, q}^f=\frac{1}{2\pi i}\, \frac{f(z) \, \chi(z) \, d z}{(z-p )\, (z-q)} \(\chi(p)-\chi(q)\). 
$$
Any point  $p \in \CP$ can be viewed as a  function on pairs,  $p: (\CP, \chi) \rightarrow \chi(p)$.   

For any two points $p$ and $q$ which are away from the poles of some $\chi_0$ we define the Poisson bracket in a  vicinity of the pair $(\CP, \chi_0)$  by the formula 
\beq\label{pb0}
\{\chi(p), \chi(q)\}_{\omega^f}\equiv \sum_{m=1}^{N}\;\,  \int\limits_{\overset{\curvearrowright}{O}_m} \omega_{p\, q}^f , 
\eeq
where $O_m$ is a small circle enclosiing the pole $z_m$ and traversed clockwise. 
Later we will prove that  (\ref{pb0}) defines a genuine Poisson bracket, {\it i.e.} it satisfies the Jacobi identity
$$
\{\{\chi(p), \chi(q)\},\chi(r)\}+\{\{\chi(q), \chi(r)\},\chi(p)\}+
\{\{\chi(r), \chi(p)\},\chi(q)\} =0,
$$

\subsection{Relation to the  Yang-Baxter equation.}
By Cauchy theorem the integral  in (\ref{pb0}) can be computed using residues 
$$
\{\chi(p), \chi(q)\}_{\omega^f}= \underset{p}{\res}\;  \omega_{p\, q}^f + \underset{q}{\res}\; \omega_{p\, q}^f + \underset{\infty}{\res} \; \omega_{p\, q}^f.
$$
There are two important cases when the residue at infinity vanishes. 

When $f(z)=1$, we obtain 
\beq\label{rsyb}
\{\chi(p), \chi(q)\}_{\omega^1}=\frac{(\chi(p)-\chi(q))^2}{p-q}. 
\eeq
This formula is associated with the rational solution of the Yang-Baxter equation. 
The rational solution of the Yang-Baxter equation describes  quadratic algebra, see \cite{GF}, with two generators\footnote{Here $\S(z)$ and $\D(z)$ are arbitrary functions of the parameter $z$.}  $\S(z)$ and $\D(z)$ such that
\bay
\{\S(q), \S(p)\}_{\omega^1}&=&\{\D(q), \D(p)\}_{\omega^1}=0,\label{num}\\
\{\S(q), \D(p)\}_{\omega^1}&=&\frac{\S(q) \D(p)- \S(p) \D(q) }{q-p}.\label{crnum}
\ey
These relations imply (\ref{rsyb}) for the function $\chi(z)=\S(z)/\D(z)$. It is interesting to note that 
relation (\ref{rsyb}) together with (\ref{num}) implies (\ref{crnum}). Therefore,  (\ref{rsyb}) can be viewed as an equivalent form of  rational solution of the YB equation. 

In the second case  $f(z)=z$, we have
\beq\label{tsyb}
\{\chi(p), \chi(q)\}_{\omega^z}=\frac{(p\, \chi(p)-q \chi(q))}{p-q} \times (\chi(p)-\chi(q)). 
\eeq
This case is connected with the trigonometric solution of the Y-B equation. The quadratic algebra associated with this solution is 
\bay
\{\S(q), \S(p)\}_{\omega^z}&=&\{\D(q), \D(p)\}_{\omega^z}=0,\label{nnum}\\
\{\S(q), \D(p)\}_{\omega^z}&=&\frac{1}{q-p}\[\frac{q+p}{2} \S(q) \D(p)- p \, \S(p) \D(q) \].\label{ncrnum}
\ey
These relations imply (\ref{tsyb}). Relation (\ref{tsyb}) together with (\ref{nnum}) implies (\ref{ncrnum}). Therefore (\ref{tsyb}) is an equivalent way to write a trigonometric solution of the Y-B equation. 
 
In the theory of completely integrable systems the role of  functions $\S$ and $\D$  is played by  the entries of the monodromy matrix of the associated spectral problem, \cite{FT}.  

It is interesting to look at the simplest case $f(z)=z^2$ for which the residue at infinity does not vanish. As before we compute 
$$
\{\chi(p), \chi(q)\}_{\omega^{z^2}}=\frac{(p^2 \, \chi(p)-q^2 \chi(q))}{p-q} \times (\chi(p)-\chi(q)) +c_0, 
$$
and 
$$
\{ \chi(q), c_0\}_{\omega^{z^2}}=q^2 \chi^2(q)+ (c_0 q + c _1) \chi(q). 
$$  
We see that  quadratic algebra is not closed and it has to be extended by  coefficients at infinity.

\subsection{Canonical coordinates.} 
Generically, 
parameters $z_1,\hdots,z_N$ and $\rho_1,\hdots,\rho_N$ play the role of coordinates on the space of pairs.  
\begin{thm} 
The Poisson bracket  (\ref{pb0})  in $z-\rho$ coordinates has the form 
\begin{eqnarray}
\{\rho_k,\rho_n\}&=&\frac{(f(z_k) +f(z_n)) \rho_k \,\rho_n}{ z_n-z_k}(1-\delta_k^n),\label{rr}\\
\{\rho_k,z_n\}&=& \rho_k  f(z_n) \delta_k^n, \label{rl}\\
\{z_k,z_n\}&=& 0 \label{ll}.
\end{eqnarray}
\end{thm}

\noi
{\it Proof. } The proof resembles the proof of Theorem 2 in \cite{V1}, but is somewhat simpler due to  a general nature of our approach. 

One can represent $\rho$'s and $z$'s using contour integrals
$$
\rho_k=-\frac{1}{2\pi i} \int\limits_{\overset{\curvearrowleft}{O_k}} \chi(\zeta) d\zeta,  \quad\quad 
\rho_k z_k= -\frac{1}{2\pi i} \int\limits_{\overset{\curvearrowleft}{O_k}}\zeta \chi(\zeta) d\zeta.
$$
Therefore, for $k\neq n$ we have
\bey
\{\rho_k,\rho_n\}&=& \frac{1}{(2 \pi i)^2}  \int\limits_{\overset{\curvearrowleft}{O_k}} \int\limits_{\overset{\curvearrowleft}{O_n}} 
\{\chi(\zeta), \chi(\eta)\}\, d\zeta\, d\eta \\
&=& \frac{1}{(2 \pi i)^2}  \int\limits_{\overset{\curvearrowleft}{O_k}} \int\limits_{\overset{\curvearrowleft}{O_n}} 
\, d\zeta\, d\eta \; \frac{1}{2\pi i}  \int\limits_{\sum_{m}\overset{\curvearrowright}{O}_m} \frac{f(z) \, \chi(z) \, d z}{(z-\zeta )\, (z-\eta)} \(\chi(\zeta)-\chi(\eta)\) \\
&=& \frac{1}{(2 \pi i)^2}  \int\limits_{\overset{\curvearrowleft}{O_k}} \int\limits_{\overset{\curvearrowleft}{O_n}} 
\, d\zeta\, d\eta \; \frac{1}{2\pi i}  \int\limits_{\sum_{m}\overset{\curvearrowright}{O}_m} \frac{f(z) \, \chi(z) \, d z}{(z-\zeta )\, (z-\eta)}\; \chi(\zeta) \\
&-& \frac{1}{(2 \pi i)^2}  \int\limits_{\overset{\curvearrowleft}{O_k}} \int\limits_{\overset{\curvearrowleft}{O_n}} 
\, d\zeta\, d\eta \; \frac{1}{2\pi i}  \int\limits_{\sum_{m}\overset{\curvearrowright}{O}_m} \frac{f(z) \, \chi(z) \, d z}{(z-\zeta )\, (z-\eta)} \;\chi(\eta) \\
&=& A-B.
\eey
The contours are given in Figure 2.

\begin{figure}[ht]
\includegraphics[width=0.50\textwidth]{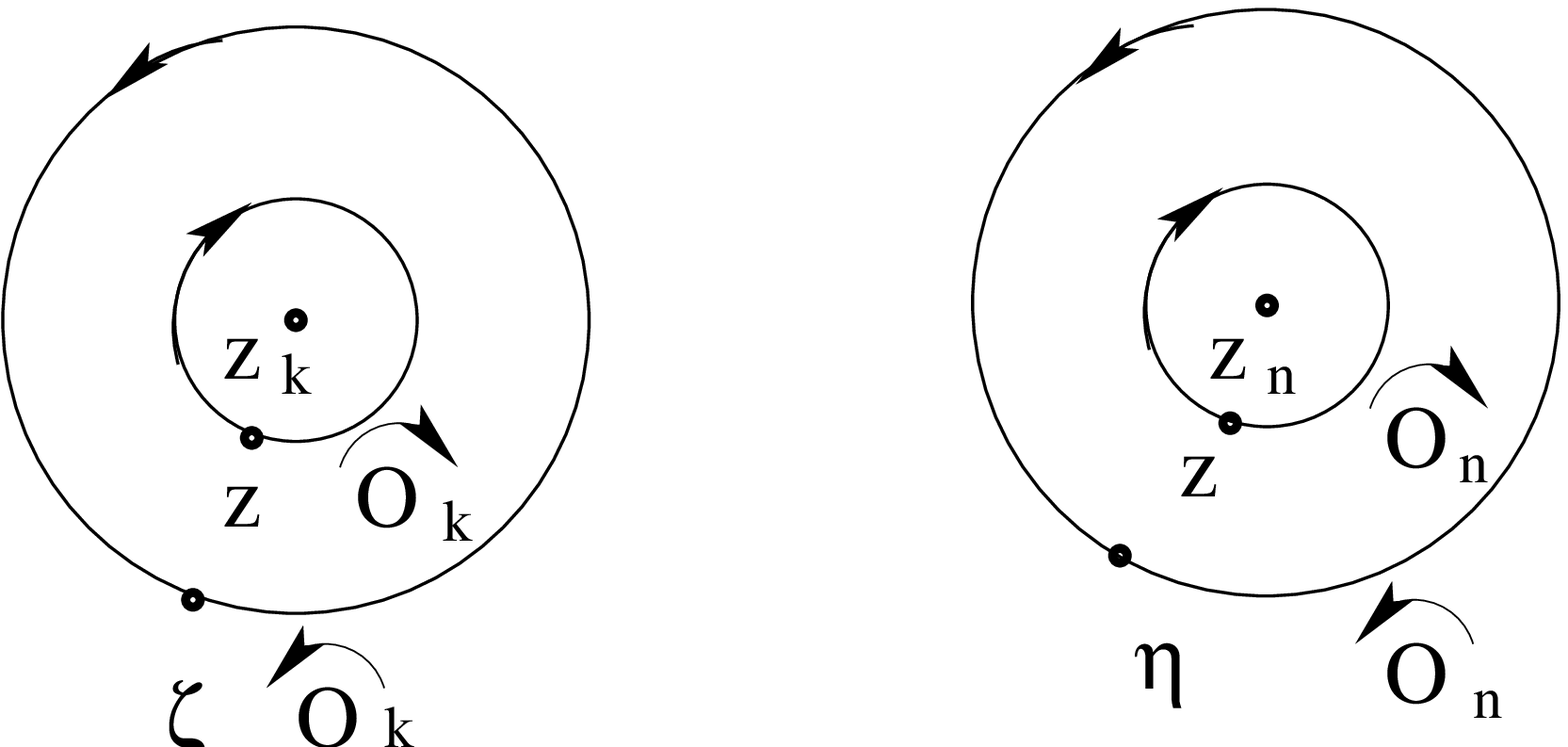}
\caption{Contours for $k\neq n$.}
\end{figure}

\noi
Using Fubini's theorem 
\bey
A&=&\frac{1}{2\pi i}  \int\limits_{\sum_{m}\overset{\curvearrowright}{O}_m} dz\,  f(z)\,  \chi(z)\;  
\times \frac{1}{2 \pi i}  \int\limits_{\overset{\curvearrowleft}{O_k}} \frac{ \chi(\zeta) d\zeta}{z-\zeta} 
\times \frac{1}{2 \pi i}  \int\limits_{\overset{\curvearrowleft}{O_n}} \frac{  d\eta}{z-\eta}\\
&=&\frac{1}{2\pi i}  \int\limits_{\overset{\curvearrowright}{O}_n} dz\,  f(z)\,  \chi(z)\;  
 \times \frac{ - \rho_k }{z-z_k} \times (-1)
= \frac{f(z_n) \rho_n \rho_k }{z_n- z_k}.  
\eey
Similarly we have 
$$
B= \frac{f(z_k) \rho_n \rho_k }{z_k- z_n} 
$$
and this implies (\ref{rr}).

To derive (\ref{rl}), for $k\neq n$ we compute
$$
\{z_k  \rho_k, \rho_n\} =  \frac{(f(z_k) +f(z_n)) z_k   \rho_k \,\rho_n}{ z_n-z_k}. 
$$ 
Onthe other hand, using Leibnitz rule 
$$
\{ z_k \rho_k, \rho_n\} = \rho_k \{z_k ,\rho_n\} + z_k \{ \rho_k,\rho_n\}. 
$$
This implies $ \{\rho_k ,z_n\}=0$  for $k\neq n$. 
Derivation is more complicated in the case $k=n$:  
\bey
\{z_n \rho_n,\rho_n\}&=& \frac{1}{(2 \pi i)^2}  \int\limits_{\overset{\curvearrowleft}{O_n}} \int\limits_{\overset{\curvearrowleft}{O_n}} 
\zeta \{\chi(\zeta), \chi(\eta)\}\, d\zeta\, d\eta \\
&=& \frac{1}{(2 \pi i)^2}  \int\limits_{\overset{\curvearrowleft}{O_n}} \int\limits_{\overset{\curvearrowleft}{O_n}} 
\zeta\, d\zeta\, d\eta \; \frac{1}{2\pi i}  \int\limits_{\sum_{m}\overset{\curvearrowright}{O}_m} \frac{f(z) \, \chi(z) \, d z}{(z-\zeta )\, (z-\eta)} \(\chi(\zeta)-\chi(\eta)\) \\
&=& \frac{1}{(2 \pi i)^2}  \int\limits_{\overset{\curvearrowleft}{O_n}} \int\limits_{\overset{\curvearrowleft}{O_n}} 
\zeta \, d\zeta\, d\eta \; \frac{1}{2\pi i}  \int\limits_{\sum_{m}\overset{\curvearrowright}{O}_m} \frac{f(z) \, \chi(z) \, d z}{(z-\zeta )\, (z-\eta)}\; \chi(\zeta) \\
&-& \frac{1}{(2 \pi i)^2}  \int\limits_{\overset{\curvearrowleft}{O_n}} \int\limits_{\overset{\curvearrowleft}{O_n}} 
\zeta \, d\zeta\, d\eta \; \frac{1}{2\pi i}  \int\limits_{\sum_{m}\overset{\curvearrowright}{O}_m} \frac{f(z) \, \chi(z) \, d z}{(z-\zeta )\, (z-\eta)} \;\chi(\eta) \\
&=& A-B.
\eey
The contours are given in Figure 3. 

\noi
Using Fubini's theorem
\bey
A&=&\frac{1}{2\pi i}  \int\limits_{\overset{\curvearrowleft}{O}_n} d \zeta \,\zeta  \,  \chi(\zeta)\;  
\times \frac{1}{2 \pi i}  \int\limits_{\sum_{m} {\overset{\curvearrowright}O_m}} 
\frac{ f(z) \chi(z) d z}{z-\zeta} 
\times \frac{1}{2 \pi i}  \int\limits_{\overset{\curvearrowleft}{O}_n} \frac{  d\eta}{z-\eta}\\
&=&\frac{1}{2\pi i}  \int\limits_{\overset{\curvearrowleft}{O}_n} d \zeta \,\zeta  \,  \chi(\zeta)\;  
\times \frac{1}{2 \pi i}  \int\limits_{ {\overset{\curvearrowright}O_n}} 
\frac{ f(z) \chi(z) d z}{z-\zeta} \times (-1)\\
&=&\frac{1}{2\pi i}  \int\limits_{\overset{\curvearrowleft}{O}_n} d \zeta \,(\zeta-z_n+z_n) \,  \chi(\zeta)\;  
\times \frac{ f(z_n) \rho_n}{z_n-\zeta} \times (-1)\\
\eey
\bey
&=&-\rho^2_n f(z_n) - z_n \rho_n f(z_n) \frac{1}{2\pi i} \int\limits_{\overset{\curvearrowleft} {O}_n} \frac{\chi(\zeta) d\zeta}{z_n-\zeta}.  
\eey

Similarly,
\bey
B&=&\frac{1}{2\pi i}  \int\limits_{\overset{\curvearrowleft}{O}_n} d \eta   \,  \chi(\eta)\;  
\times \frac{1}{2 \pi i}  \int\limits_{\sum_{m} {\overset{\curvearrowright}O_m}} 
\frac{ f(z) \chi(z) d z}{z-\eta} 
\times \frac{1}{2 \pi i}  \int\limits_{\overset{\curvearrowleft}{O}_n} \frac{ \zeta d\zeta}{z-\zeta}\\
&=&\frac{1}{2\pi i}  \int\limits_{\overset{\curvearrowleft}{O}_n} d  \eta   \,  \chi(\eta)\;  
\times \frac{1}{2 \pi i}  \int\limits_{ {\overset{\curvearrowright}O_n}} 
\frac{ f(z) \chi(z) d z}{z-\eta} \times (-z)\\
&=&\frac{1}{2\pi i}  \int\limits_{\overset{\curvearrowleft}{O}_n} d \eta \,   \chi(\eta)\;  
\times \frac{ - z_n f(z_n) \rho_n}{z_n-\eta} \\
&=& - z_n \rho_n f(z_n) \frac{1}{2\pi i} \int\limits_{\overset{\curvearrowleft} {O}_n} \frac{\chi(\eta) d\eta}{z_n-\eta}.  
\eey
Therefore,
$$
\{z_n \rho_n, \rho_n\} =-\rho_n^2 f(z_n),
$$
and this implies (\ref{ll}). 

\begin{figure}[ht]
\includegraphics[width=0.40\textwidth]{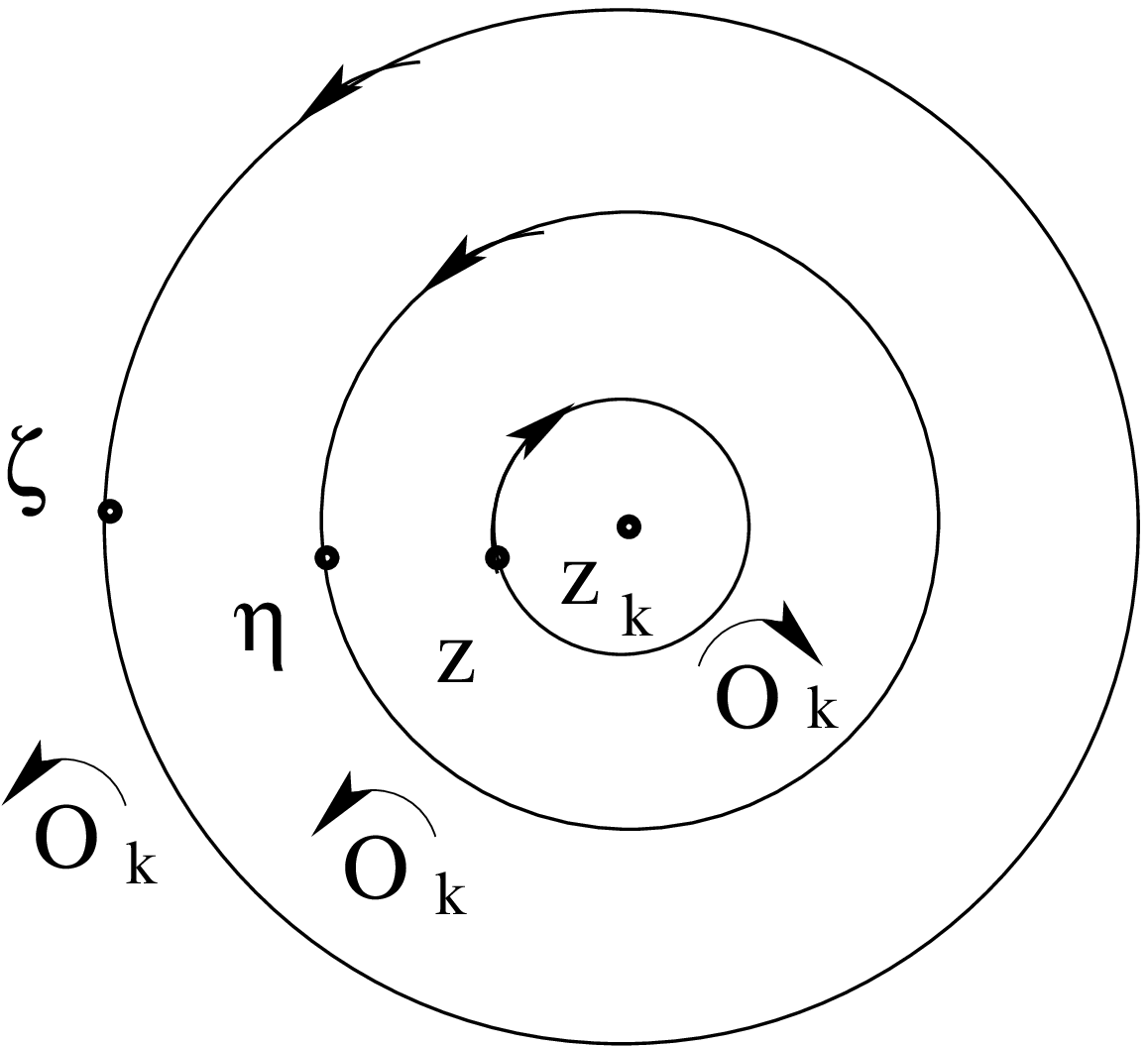}
\caption{Contours for $k= n$.}
\end{figure}

To obtain  (\ref{rr}), for $k\neq n$, as before  we compute
$$
\{ z_k \rho_k, z_n \rho_n\} =  \frac{(f(z_k) +f(z_n)) z_k z_n   \rho_k \,\rho_n}{ z_n-z_k}. 
$$ 
On the other hand,
$$
\{z_k  \rho_k,z_n \rho_n\} =  z_k z_n \{ \rho_k,\rho_n\} +z_k \rho_n \{\rho_k ,z_n\} + \rho_k z_n \{z_k ,\rho_n\} + \rho_k \rho_n \{ z_k,z_n\} . 
$$
This together with (\ref{rl}) implies (\ref{ll}).

\qed

\noi
One can easily recompute the bracket (\ref{pb0}) in terms of the coordinates
$$
z_1,\hdots,z_N,\quad\quad\quad S(z_1),\hdots, S(z_N).
$$
Namely, using $S(z_k)= - \Delta'(z_k) \rho_k$ we  obtain 
$$
\{S(z_k),z_n\} = f(z_n) S(z_k) \delta_k^n,
$$
and all other brackets vanish
$$
\{S(z_k),S(z_n)\} =\{z_k,z_n\} =0. 
$$
Note that this result implies the  Jacobi identity  for the bracket and compatibility of the brackets with different $f$'s.

\section{The Liouville correspondence and spectral theory of the string}
\subsection{Basic constructions.}
We consider the  Camassa--Holm  equation  (\ref{chh})
in the  space  $\MM$ of all smooth nonnegative functions decaying fast enough  at infinity so that (\ref{dec}) holds.
We also need $\MM_0 \in \MM$ -  the subspace of functions which vanish  far enough  to the left.  
$\MM$ and $\MM_0$ are invariant under the CH flow, \cite{CM}.

The CH equation  is a compatibility condition between 
\beq\label{slsp}
D^2 f-{1\over 4}\,  f + \l m f=0  
\eeq
and 
$$
f^{\bullet} =-\(v+{1\over 2\l}\) Df +{1\over 2} (Dv) f;
$$
 i.e., $(D^2 f)^{\bullet} =D^2(f^{\bullet})$ is equivalent to (\ref{chh}).

The standard Liouville's  transformation 
$$
x \rightarrow \xi(x)=2 \tanh x/2,\qquad\qquad
f(x) \rightarrow f(\xi)={f(x)\over \cosh x/2}
$$ 
converts  (\ref{slsp})   into the string  spectral problem     
\beq\label{ssp}
f''+ \l g f =0,    
\eeq
with $g(\xi)=m(x)\cosh^4 x/2$ and $-2 \leq \xi \leq +2$. 
The transformation changes the length element  by the rule 
$$
dx\longrightarrow d\xi = dx \, \J(\xi), \qquad\qquad \text{with} 
\qquad \J(\xi)= 1-{\xi^2\over 4}.
$$
Under condition (\ref{dec}), the string is regular,  {\it i.e.}, its mass is finite: 
$$
\int\limits_{-2}^{+2} g(\xi) \, d\xi=\int\limits_{-\infty}^{+\infty} 
m(x) \cosh^2 x/2 \, dx < \infty.
$$
Evidently, for initial data from $\MM_0$ there is an interval of length $l$  where the potential vanishes:  
$g(\xi)=0,\quad\xi \in [-2,-2+l]$. 
The  transformation  reduces  problem (\ref{slsp}) with two singular ends to the regular string  
on  finite interval. 

The  spectral theory of a string with nonnegative 
mass  was constructed by M.G. Krein in the 1950's and is presented in \cite{KK2}, see also \cite{GK,DM}.
  
To formulate the results we need to  introduce two solutions $\varphi(\xi,\l)$ and $\psi(\xi,\l)$ of 
the eigenvalue problem (\ref{ssp}) with standard normalization (see (\ref{init_val})).
The functions $\varphi(2,\l)$  and  $\psi(2,\l)$  can be written in the form
\beq\label{prod}
\varphi(2,\l)=\prod_{k=1}^{\infty}\(1-{\l\over \m_k}\),\qquad\qquad \psi(2,\l)=4\prod_{k=1}^{\infty}\(1-{\l\over \l_k}\). 
\eeq
Their roots interlace each other
$$
0<  \m_1 <\l_1 < \m_2 < \l_2 <\hdots.
$$

For the string $\S_0$ with  fixed left  and   right ends   according to the general theory  
\beq\label{gggg}
E_0(\l)=-{\varphi(2,\l)\over \psi(2,\l)}= -{1\over 4}+\sum_{k=1}^{\infty} \(\frac{1}{\l_k-\l}-\frac{1}{\l_k}\)\rho_k.  
\eeq
The poles of $E_0(\l)$ are zeros of $\psi(2,\l)$  and  they do not move under the CH flow. 
Introducing
\beq\label{one}
z_k=-{1\over \l_k},\qquad\qquad \rho_k'={\rho_k\over \l_k^2},   
\eeq
we have the identity
\beq\label{two}
{\rho_k\over \l_k-\l}={\rho_k'\over z_k-z}+ {\rho_k\over \l_k}.  
\eeq
Therefore   for the function $E_0$  we obtain
\beq\label{three}
E_0(z)= -{1\over 4} + \sum_{k=1}^{\infty} {\rho_k'\over z_k-z}.
\eeq
In this form it is suitable for our purposes. 


\subsection{The first and the second  analytic Poisson brackets for the  Weyl function} 
\noi
Very little have to be added now in order to introduce a family of compatible Poisson brackets. The pair 
$(\Gamma, E_0)$ provides a parametrization of the phase space $\MM$, see \cite{V0}. We will write Poisson brackets in terms of the pair. 

Consider a differential $\omega_{p \, q}^{f}$ on the lower component of the spectral curve $\Gamma_-$ which depends on the  
entire function $f$ and two points $p$ and $q$ 
$$
\omega_{p\, q}^f=\frac{1}{2\pi i}\, \frac{f(z) \, E_0(z) \, d z}{(z-p )\, (z-q)} \(E_0(p)-E_0(q)\). 
$$
Any point  $p \in \Gamma_-$ can be considered as a  function on pairs $p: (\Gamma, E_0 ) \rightarrow E_0(p)$.   
For any two points $p$ and $q$ which are away from the poles of some  $E_0$ we define the Poisson bracket in a  
vicinity of the pair $(\Gamma, E_0)$  by the formula 
\beq\label{rreerr}
\{E_0(p), E_0(q)\}_{\omega^f}=\sum_{m}\;\,  \int\limits_{\overset{\curvearrowright}{O}_m} \omega_{p\, q}^f, 
\eeq
where the circles $O_m$ are traversed clockwise and surround points $z_m$. The sum over all small contours converges  since its  may be replaced by one integral over a large contour  surrounding all poles $z_k$.

Now we need the following

\noi
\begin{thm}\label{ddd} \cite{V0}.  Let $N(\zeta)=N_0$ be a real constant, possibly infinity. 
Then,
\beq\label{ppo}
\{\En(\l),\En(\m)\}_{J_0}={\l\m\over \l-\m}\(\En(\l)-\En(\m)\)^2.  
\eeq
\end{thm}

\noi
If one changes the spectral parameter by the rule 
$$
p = -{1\over \l},\qquad\qquad\qquad q = 
-{1\over \m},
$$ 
then the  formula (\ref{ppo}) becomes
$$
\{\En(p),\En(q)\}_{J_0}={\(\En(p)-\En(q)\)^2\over p-q}.
$$
Therefore, comparing this with  (\ref{rsyb}) we see that the Poisson bracket $J_0=mD+Dm$ corresponds to the choice $f(z)=1$ in the formula (\ref{rreerr}).  

It can be proved either by similar direct calculation or by using the recurrence relation \ref{recc} that the second Poisson bracket $J_1=D-D^3$ corresponds 
to the choice $f(z)=z$ in \ref{rreerr}. 

The formula (\ref{rreerr}) defines the Poisson bracket  for {\it any} choice of an entire function $f(z)$, 
though its explicit expression in terms of the phase space  $\MM$ is not known to us. It is not even known when the corresponding Poisson tensor on  $\MM$ is  local.

\newpage

\noindent
Michael Gekhtman
\newline
Department of Mathematics
\newline
University of Notre Dame
\newline
Notre Dame, IN 
\newline
USA
\vskip 0.1in
\noindent
mgekhtma@nd.edu

\vskip 0.5 in
\noindent
Kirill Vaninsky
\newline
Department of Mathematics
\newline
Michigan State University
\newline
East Lansing, MI 48824
\newline
USA
\vskip 0.1in
\noindent
vaninsky@math.msu.edu

\end{document}

%% file: definitions.tex
\newcommand{\noi}{\noindent}

\def\res{\text{ res}}

\newcommand{\CP}{{\mathbb C}{\mathbb  P}}

\newcommand{\J}{{\mathcal J}}

\newcommand{\HH}{{\mathcal H}}
\newcommand{\MM}{{\mathcal M}}

\def\En{E_{N_0}}

\def\S{\mathcal S}

\def\[{\left[}
\def\]{\right]}
\def\({\left(}
\def\){\right)}

\def\l{\lambda}

\def\m{\mu}

\def\D{\Delta}

\def\12{{1\over 2}}

\newcommand{\p}{\partial}
\newcommand{\eeq}{\end{equation}}
\newcommand{\beq}{\begin{equation}}
\newcommand{\bay}{\begin{eqnarray}}
\newcommand{\ey}{\end{eqnarray}}
\newcommand{\bey}{\begin{eqnarray*}}
\newcommand{\eey}{\end{eqnarray*}}

\newcommand{\Rat}{\operatorname{Rat}}

\newtheorem{thm}{\hspace{\parindent}Theorem}[section]